\documentclass{emulateapj}
\usepackage{textcomp}
\usepackage{mathtext}
\usepackage{amssymb}
\usepackage{rotating}

\newcommand{\fermi}{\textit{Fermi}}
\newcommand{\gr}{$\gamma$-ray}
\newcommand{\gro}{GRO J1008$-$57}

\begin{document}

\title{The likely {\it Fermi} detection of the Be X-ray binary GRO J1008$-$57}
%%%hint for $\gamma$-ray production in neutron-star X-ray binaries}

\author{Yi Xing \& Zhongxiang Wang}

\affil{Key Laboratory for Research in Galaxies and Cosmology, Shanghai Astronomical Observatory, Chinese Academy of Sciences,
80 Nandan Road, Shanghai 200030, China}

\begin{abstract}
In our search for $\gamma$-ray emission from Be X-ray binaries from analysis
of the data obtained with the Large Area Telescope (LAT) on board the
{\it Fermi Gamma-Ray Space Telescope}, we find likely detection of 
GRO~J1008$-$57. The binary has an orbital period of 249.48 days, and
it is only significantly detected in its orbital phase
0.8--0.9 ($> 4\sigma$). Further
light curve analysis indicates that the detection is probably largely due
to an emitting event in one orbital cycle around year 2012--2013, following
a giant X-ray outburst of the source.
This property of having occasional $\gamma$-ray emitting events is similar
to that seen in another high-mass X-ray binary 4U 1036$-$56. However, models
considering possible $\gamma$-ray emission from an accreting neutron star have 
difficulty in explaining the observed $\sim 10^{34}$\,erg\,s$^{-1}$ luminosity
of the source, unless the distance was largely over-estimated. Further 
observational studies are required, in order to more clearly 
establish the high-energy emission properties of \gro\ or similar high-mass 
X-ray binaries and find clues for understanding
how $\gamma$-ray emission is produced from them.
%%%The combination of the stellar rotation, magnetic field, and mass accretion of the neutron star (a transient X-ray pulsar with a spin period of 93.5 s) makes GRO~J1008$-$57 a system in which its propeller phase can be switched on in the quiescent state when its X-ray luminosity is lower than $\sim 10^{34}$ erg\,s$^{-1}$.  Considering these features, we suggest that occasional $\gamma$-ray emitting events occur in GRO~J1008$-$57. Particularly after a giant outburst, which probably would exhaust large amount of disk matter and induce the propeller phase in the following orbital phase, then similar to the transitional pulsar binaries (with PSR J1023+0038 as a prototype), the neutron star's propellering processes or even its partial pulsar wind would result in the observed $\gamma$-ray emission. If this is confirmed, GRO~J1008$-$57 may represent a type of transient $\gamma$-ray sources among neutron-star XRBs.

\end{abstract}


\section{Introduction}

X-ray binaries (XRBs) contain a compact primary, either a black hole
or a neutron star, and when the companion is an early-type (O/B) massive star,
such XRBs are classified as high-mass X-ray binaries (HMXBs). 
Generally the compact star in an HMXB system accretes from matter 
carried in the stellar wind of the companion and in most cases 
the accretion rates are low (e.g., \citealt{do73,lvp76,wal+15}). 
High X-ray luminosities (10$^{35}$--10$^{40}$ erg s$^{-1}$) can be 
observed when the compact star is 
interacting with the dense part of the stellar wind of a Be star companion. 
Be XRBs account for a majority of the known HMXBs 
($\sim$50\%; \citealt{wal+15}), in most of which the compact stars are 
magnetized neutron stars \citep{lvv06,lt09}. 

There are three Be XRBs, 
PSR B1259$-$63/LS 2883 \citep{aha+09,tam+11,abd+11}, 
LS I $+$61\textdegree303 \citep{abd+09,had2012}, 
and HESS 0632$+$057 \citep{hin+09,bon+11,li+17_0632}, have been seen to 
emit photons in GeV--TeV \gr\ band. They are also classified as \gr\ binaries 
as their spectral energy distributions peak in \gr\ energies \citep{dub13}.
$\gamma$-ray emission from them is orbitally modulated, and
interestingly a superorbital periodicity is seen 
in LS I $+$61\textdegree303 (\citealt{ack+13,ahn+16} and references therein). 
PSR B1259$-$63/LS 2883 is the only \gr\ binary with a known
$\sim$47.8 ms radio pulsar \citep{joh+92}. The pulsar moves around 
the Be-type companion with a long orbital period of $\sim$3.4 years 
and a high eccentricity of 0.87 \citep{joh+94}, and during the periastron 
passage a strong flare particularly at high-energies of X-ray to \gr\ is 
induced due to the close
interaction between the winds of the two stars (e.g., \citealt{che+14,tam+15}; 
see also \citealt{xwt16}).
Recently another pulsar system PSR 2032$+$4127/MT91 213 has also been 
found to be a candidate \gr\ binary system with a pulsar moving around 
a Be-type companion in a $\sim$50 year long orbit \citep{lyn+15,ho+17}.

Currently there are only five confirmed \gr\ binaries in the Galaxy (including
the above three). While the limited number may provide constraints on
binary evolution and high-energy physical processes
(e.g., \citealt{mv89,dub13}), searches for more members of this class or
related sources with \gr\ emission will help improve our understanding.
Be XRBs are good candidates for such searches because of the existence of
the circumstellar disks, providing an environment for different possible
interactions with the neutron star primaries. For example,
\citet{rom+01} has reported the likely detection of the variable \gr\ emission 
from a Be XRB system A~0535+26. The \gr\ emission was suggested to originate
in hadronic processes, in which hadrons accelerated from
the magnetosphere of the neutron star could impact the surrounding accretion 
disk and produce \gr\ photons via the neutral pion decay process 
(e.g., \citealt{cr91}). However, no \gr\ emission from this source was detected
during a giant X-ray outburst \citep{acc+11}.
It was also predicted that A~0535+26 would have a high neutrino 
yield \citep{anc+03}, which is not found with IceCube.
Motivated by these and taking advantage of the all-sky monitoring capability
of the Large Area Telescope (LAT) onboard the {\it Fermi Gamma-ray Space 
Telescope (Fermi)}, we conducted search for \gr\ emission from Be
XRBs. The targets were selected from those given in \citet{wal+15}. 
Considering the X-ray and \gr\ emission from this type of sources are 
generally orbitally modulated, we only included the sources with known 
orbital parameters, allowing to conduct orbital-phase resolved search.
Such search may be more sensitive due to orbitally-dependent physical processes
(see, e.g., \citealt{xwt16}).

In our search, likely detections of \gro\ in its certain orbital phase ranges
were found. This Be XRB, discovered in 1993 \citep{sto+93}, contains a spin 
period $P=93.5$\,s transient X-ray pulsar, whose pulsed emission was detectable
during the source's X-ray outbursts (e.g., \citealt{coe+07,kuh+13}). 
The binary has an orbit with a period of 249.48 days and an eccentricity of
0.68, determined from
long-term timing of the pulsar's pulsed emission \citep{coe+07,kuh+13}. 
The magnetic field of the pulsar is known to be the highest among 
the Be XRBs, likely as high as $\sim$8$\times$10$^{12}$\,G 
given the suggested cyclotron line at $\sim$88 keV \citep{shr+99}.
\gro\ exhibits type-I X-ray outbursts periodically at each periastron 
passage (e.g., \citealt{tsy+17}), due to the interaction between the neutron
star and the circumstellar disk around the Be 
companion (e.g., \citealt{rei11}), and occasionally type-II outbursts. 
The latter type of the outbursts in Be XRBs, which are much stronger 
than type-I, is believed to cause major changes in the structure of the
circumstellar disk, sometimes even leading to the disappearance of the disk
\citep{rei11}.

In this paper, we report our results obtained for \gro\ from our analysis
of the \fermi\ LAT data.
\begin{figure}
\centering
\epsscale{1.0}
\plotone{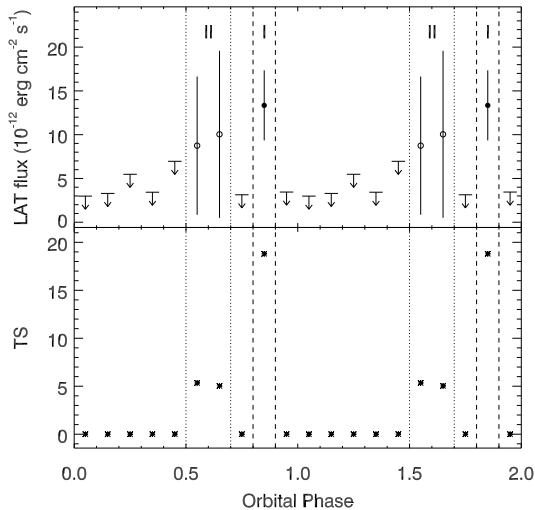}
\caption{Orbital light curve and TS curve (0.3--500 GeV) for \gro\ obtained
from \fermi\ LAT data. Flux 
points with TS greater than 9 and 5 are plotted with solid and open circles 
respectively, and the 95\% upper limits are plotted otherwise.}
\label{fig:orbital}
\end{figure}

\section{\fermi\ LAT Data Analysis and Results}
\label{sec:obs}

\subsection{LAT Data and Source Model}

LAT scans the whole sky every three hours in the energy band of 
0.1--500 GeV \citep{atw+09}. In the analysis, we selected LAT events 
from the \textit{Fermi} Pass 8 database in the time period from 
2008-08-04 15:43:36 (UTC) to 2017-10-19 00:56:35 (UTC). 
For \gro, a $\mathrm{20^{o}\times20^{o}}$ region centered at its position
was selected.  We followed the recommendations of the LAT 
team\footnote{\footnotesize http://fermi.gsfc.nasa.gov/ssc/data/analysis/scitools/}
by including the events with zenith angles less than 90 degrees (to prevent 
the Earth's limb contamination) and excluding the events with 
quality flags of `bad'.

All sources within a 20 degree region centered at the target
in the \textit{Fermi} LAT 4-year catalog \citep{3fgl15}
were included to make the source model. Spectral forms of these 
sources are provided in the catalog. Spectral parameters of the sources 
within 5 degrees from the target were set as free parameters, and 
parameters of other sources were fixed at their catalog values. 
\gro\ was included in the source model as a point source 
with power-law emission. In addition, the background Galactic 
and extragalactic diffuse emission were added in the source model 
using the spectral model gll\_iem\_v06.fits and 
file iso\_P8R2\_SOURCE\_V6\_v06.txt respectively.  
The normalizations of the diffuse components were free parameters
in the analysis.

\subsection{Source Search in Whole Data}
\label{subsec:la}

Using the LAT science tools software package {\tt v11r5p3}, we first performed 
standard binned likelihood analysis to the LAT data for \gro. 
Since below 300 MeV, the instrument response function of the LAT has
relatively large uncertainties 
and Galactic background emission is also strong, we chose events   
above 300 MeV for the likelihood analysis. The obtained Test Statistic (TS) 
value at the position of \gro\ was $\simeq$11.
The TS value at a position indicates the fit improvement for including 
a source, and is approximately the square of the detection significance 
of the source \citep{1fgl}.
Therefore we found possible detection with $>$3$\sigma$ significance.

We checked whether the \gr\ emission was due to
contamination by nearby sources which were not included in 
the \textit{Fermi} LAT 4-year catalog. A preliminary LAT 8-year point source 
list was released in early 2018, which contains nearly 2500 new sources, 
although it is not encouraged to use this list directly\footnote{\footnotesize https://fermi.gsfc.nasa.gov/ssc/data/access/lat/fl8y/}. 
In addition, the extended source templates and the Galactic and extragalactic 
diffuse emission models have not been updated accordingly. We thus
only added nearby new sources to the source model,
and re-performed the maximum likelihood analysis. 
In this analysis,  we only found TS$\simeq$7 at the position of \gro, which
is low for a possible detection.

\subsection{Orbital-phase Resolved Search}
\label{subsec:ova}

Because Be XRBs often show enhanced emission at certain orbital phase ranges,
particularly indicated by those \gr\ binaries, we searched for possible 
\gr\ emission in 10 orbital phase ranges of \gro\ 
(i.e., 0.0--0.1, ..., 0.9--1.0, with phase zero at the periastron).
Likelihood analysis to the data in each of the phase bins 
was performed. We found during phase 0.8--0.9 and 0.5--0.7 (defined as
Phase I and II, respectively), the source was possibly detected with
TS$>$9.  In Figure~\ref{fig:orbital}, its 0.3--500 GeV orbital light curve 
and TS curve in 10 orbital-phase bins are shown.
For the data points not in Phase I and II, their TS values were smaller
than 5 ($<$2$\sigma$ significance), and the 95\% (at 2$\sigma$ level) 
flux upper limits were derived.
\begin{figure*}
\centering
\includegraphics[width=0.32\textwidth]{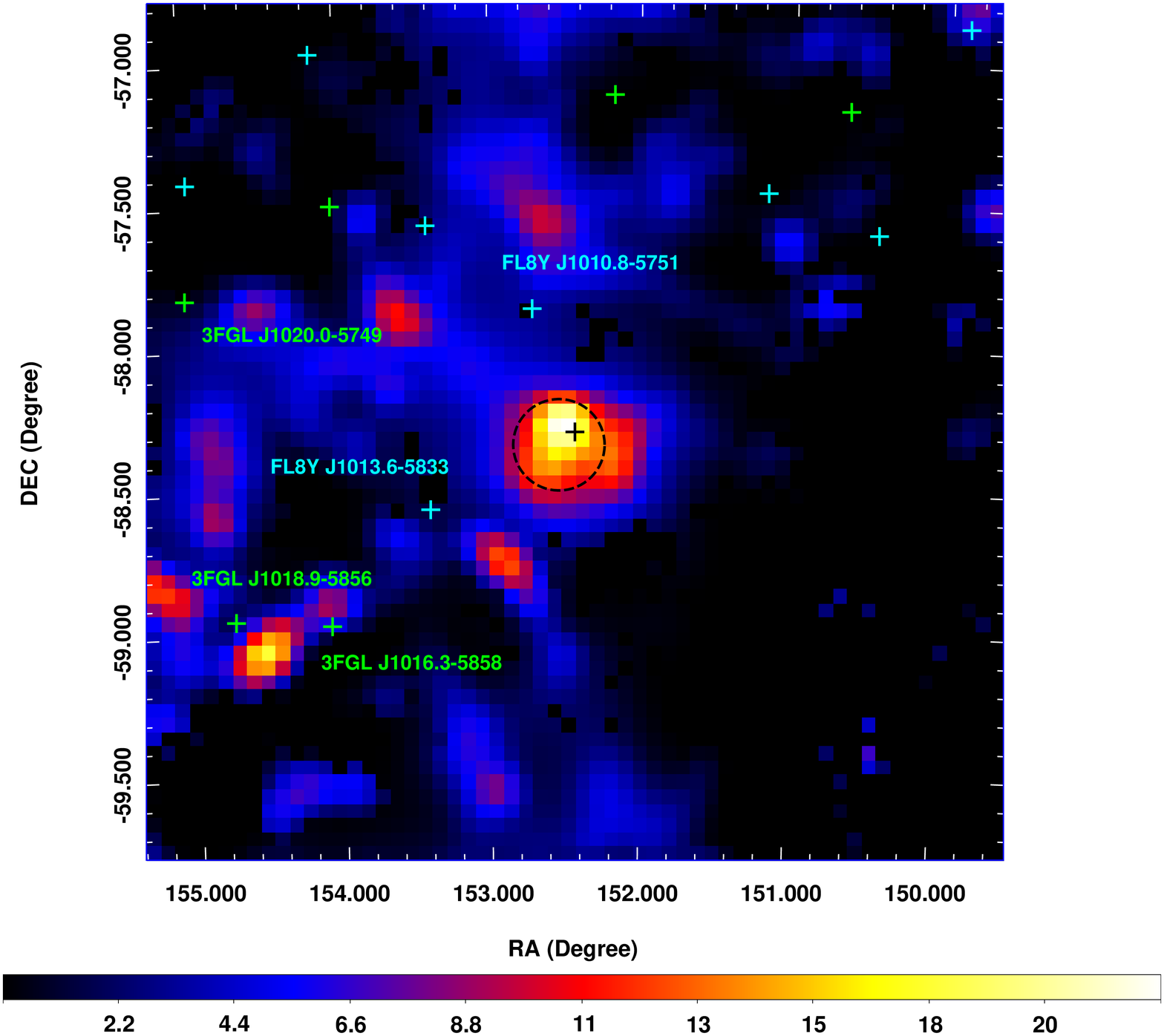}
\includegraphics[width=0.32\textwidth]{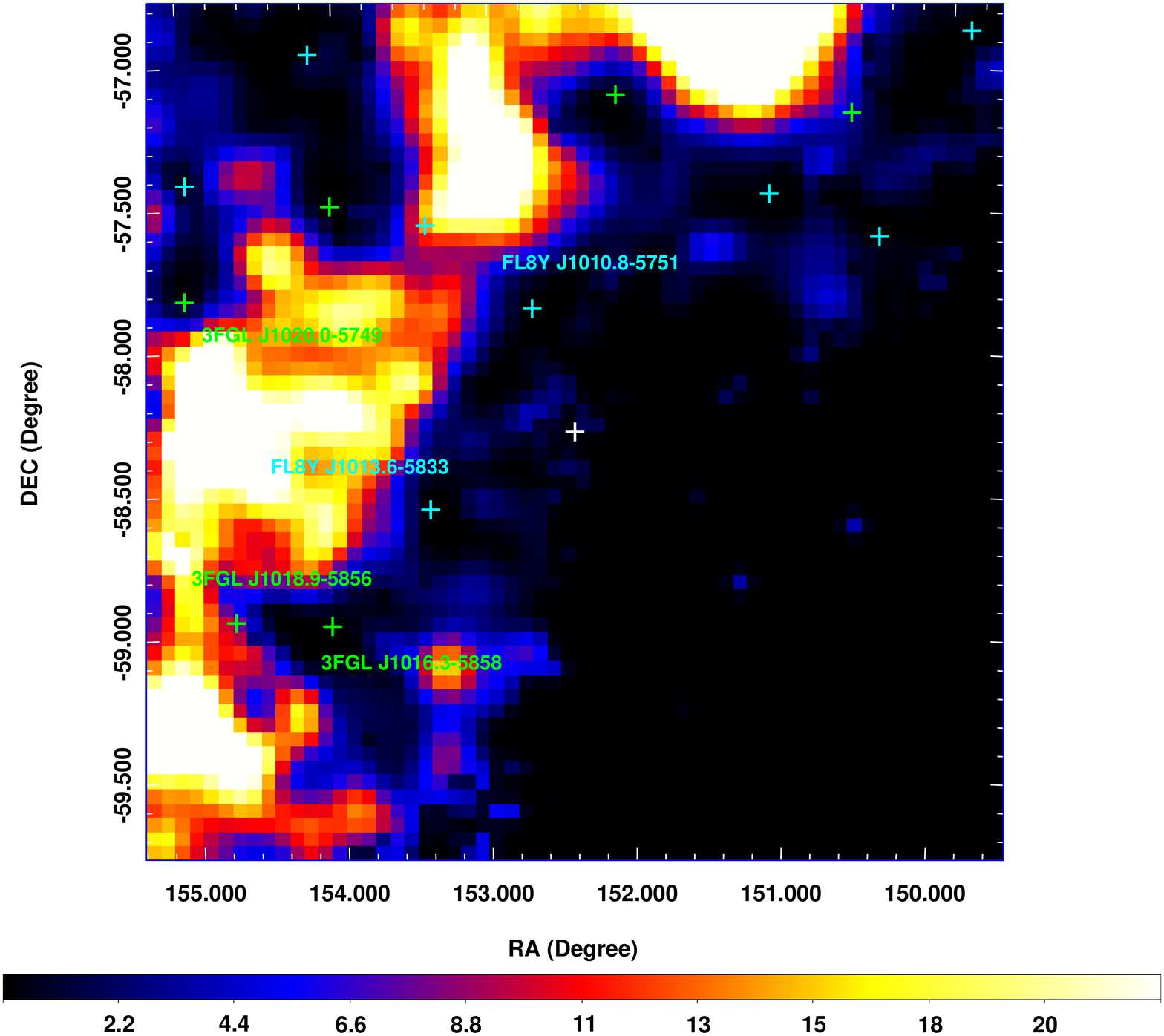}
\caption{Residual TS maps for \gro, with a size of $\mathrm{3^{o}\times3^{o}}$, 
in 0.5--39 GeV band during Phase I ({\it left}) and the phase ranges 
excluding Phase I ({\it right}).
All sources in the source model were considered and removed. 
The image scale of the maps is 0\fdg05 pixel$^{-1}$, with the color bar 
indicating the TS value range. The green and light blue pluses mark 
the positions of the catalog sources listed in the 4-year and 8-year 
\fermi\ LAT source catalogs, respectively. The black (or white) plus 
and black circle mark the position of \gro\ and the 2$\sigma$ error circle of 
the best-fit position, respectively.}
\label{fig:tsmap-fullband}
\end{figure*}

\subsubsection{Phase I}
\label{subsec:fb}
The \gr\ emission was significantly detected with a TS value of $\sim$20.
Photon index $\Gamma= 2.2\pm$0.2 and 0.3--500 GeV flux 
$F_{0.3-500}= 13\pm 4\times 10^{-12}$ erg~s$^{-1}$\,cm$^{-2}$ were 
obtained during Phase I from the maximum likelihood analysis.
We extracted the \gr\ spectrum by performing 
maximum likelihood analysis of the LAT data in 10 evenly 
divided energy bands in logarithm from 0.1--500 GeV. 
In the extraction, the spectral normalizations of the sources within 5 
degrees from \gro\ were set as free parameters, while the other parameters 
of the sources were fixed at the values obtained from the above maximum 
likelihood analysis. A point source with power-law emission was 
assumed for \gro, and the $\Gamma$ value was fixed to 2.
We kept only spectral data points with the flux values 2 times 
greater than the uncertainties, and otherwise derived 95\% 
(at 2$\sigma$ level) flux upper limits. 
The obtained flux and TS values of the spectral data points are given
in Table~\ref{tab:spectra}.
We also evaluated the systematic uncertainties of the spectral points 
induced by the uncertainties of the LAT effective area and the Galactic 
diffuse model. The former are 10\% at 100 MeV, 5\% at 500 MeV, and 20\% at 
10 GeV \citep{ran+09}, and the latter is a dominant one 
in the full LAT energy band, which was estimated by varying the normalization 
by $\pm$6\%.
The obtained systematic uncertainties are also given in 
Table~\ref{tab:spectra}.
\begin{table}
\tabletypesize{\footnotesize}
\tablecolumns{10}
\tablewidth{240pt}
\setlength{\tabcolsep}{2pt}
\caption{\fermi\ LAT flux measurements of \gro}
\label{tab:spectra}
\begin{tabular}{lccccccccccccc}
\hline
\multicolumn{2}{c}{ } &
\multicolumn{2}{c}{Phase I} &
\multicolumn{2}{c}{Phase II} \\ \hline
$E$ & Band & $F/10^{-12}$ & TS & $F/10^{-12}$ & TS \\
(GeV) & (GeV) & (erg cm$^{-2}$ s$^{-1}$) &  & (erg cm$^{-2}$ s$^{-1}$) & \\ \hline
0.15 & 0.1--0.2 & 9.1 & 0 & 6.8 & 0 \\
0.36 & 0.2--0.5 & 5.2 & 0 & 3.5 & 0 \\
0.84 & 0.5--1.3 & 5.7$\pm$2.8$\pm$1.0 & 9 & 1.1 & 0 \\
1.97 & 1.3--3.0 & 5.1 & 0 & 2.2 & 0 \\
4.62 & 3.0--7.1 & 2.4$\pm$1.2$\pm$0.1 & 6 & 2.4 & 2 \\
10.83 & 7.1--16.6 & 4 & 1 & 2.0 & 1 \\
25.37 & 16.6--38.8 & 2.58$\pm$1.28$\pm$0.01 & 7 & 2.68$\pm$1.30$\pm$0.03 & 10 \\
59.46 & 38.8--91.0 & 5.2 & 0 & 3.1 & 0 \\
139.36 & 91.0--213.3 & 10.8 & 0 & 5.3 & 0 \\
326.60 & 213.3--500.0 & 22.2 & 0 & 30.8 & 2 \\
\hline
\end{tabular}
\vskip 1mm
\footnotesize{Note: $F$ is the energy flux ($E^{2}dN/dE$). 
The first and second uncertainties on fluxes are the statistical and 
systematic ones,} and fluxes without uncertainties are 95$\%$ upper limits.
\end{table}

We also calculated the residual TS map during Phase I in the energy 
range of 0.5--39 GeV.  This energy range covers the three spectral data points 
with TS$>$4 (Table~\ref{tab:spectra}).
The TS map is shown 
in the left panel of Figure~\ref{fig:tsmap-fullband}. 
There is \gr\ emission present at the position of \gro\ with 
a TS of $\sim$22.
We ran \textit{gtfindsrc} to determine the position of the \gr\
emission, and obtained R.A.=152\fdg5, 
Decl.=$-$58\fdg3, (equinox J2000.0), with 1$\sigma$ 
nominal uncertainty of 0\fdg1.
\gro\ is 0\fdg07 from this position and within the 1$\sigma$ error circle, 
indicating the likely association.
As a comparison, the 0.5--39 GeV TS map excluding the data in Phase I 
was also calculated (shown in the right panel of 
Figure~\ref{fig:tsmap-fullband}). No \gr\ emission is present
as TS$\sim$0 at the position of \gro.
The comparison of the two TS maps support the detection of \gro\ in Phase I.

\subsubsection{Phase II}
\label{subsec:hc}

There are also marginal detections with $>$2$\sigma$ significance
during phase 0.5--0.7.  We searched for possible detection during 
this wide phase range, and found that the \gr\ emission can be 
detected with a TS of 10 ($>$3$\sigma$ significance).  
$\Gamma= 1.5\pm$0.3 and 
$F_{0.3-500}= 9\pm 7\times 10^{-12}$ erg~s$^{-1}$\,cm$^{-2}$ were obtained 
from the maximum likelihood analysis. 
We extracted the \gr\ spectrum during Phase II,
and the obtained flux and TS values of the spectral data 
points are provided in Table~\ref{tab:spectra}.
There is actually only one spectral data point possibly detected,
TS$\simeq$10 in energy band 
of 16.6--38.8~GeV. Comparing to the results obtained in Phase I,
$\Gamma$ is lower in Phase II, although the uncertainties are too large for
drawing a conclusion.

\begin{figure*}
\centering
\includegraphics[width=0.32\textwidth]{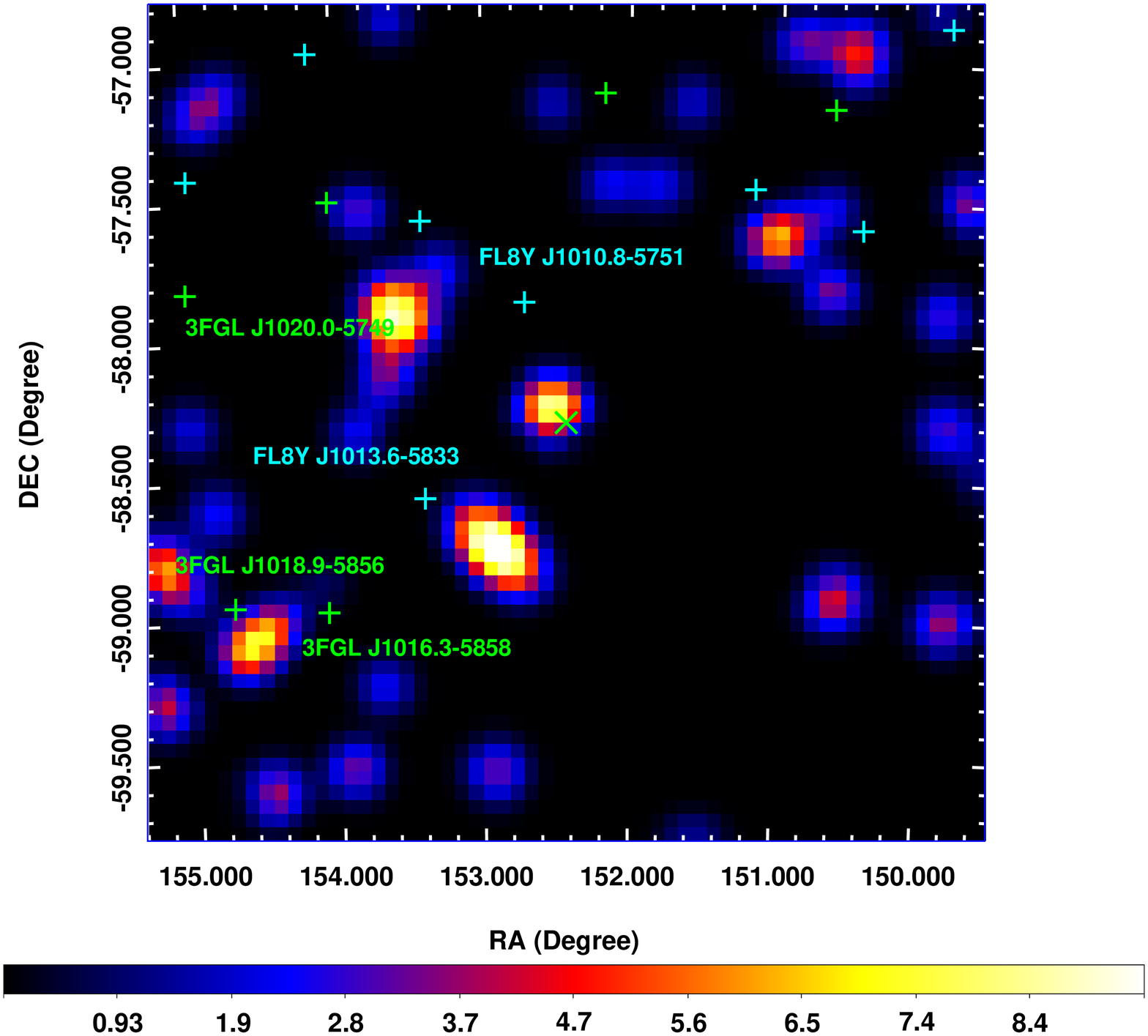}
\includegraphics[width=0.32\textwidth]{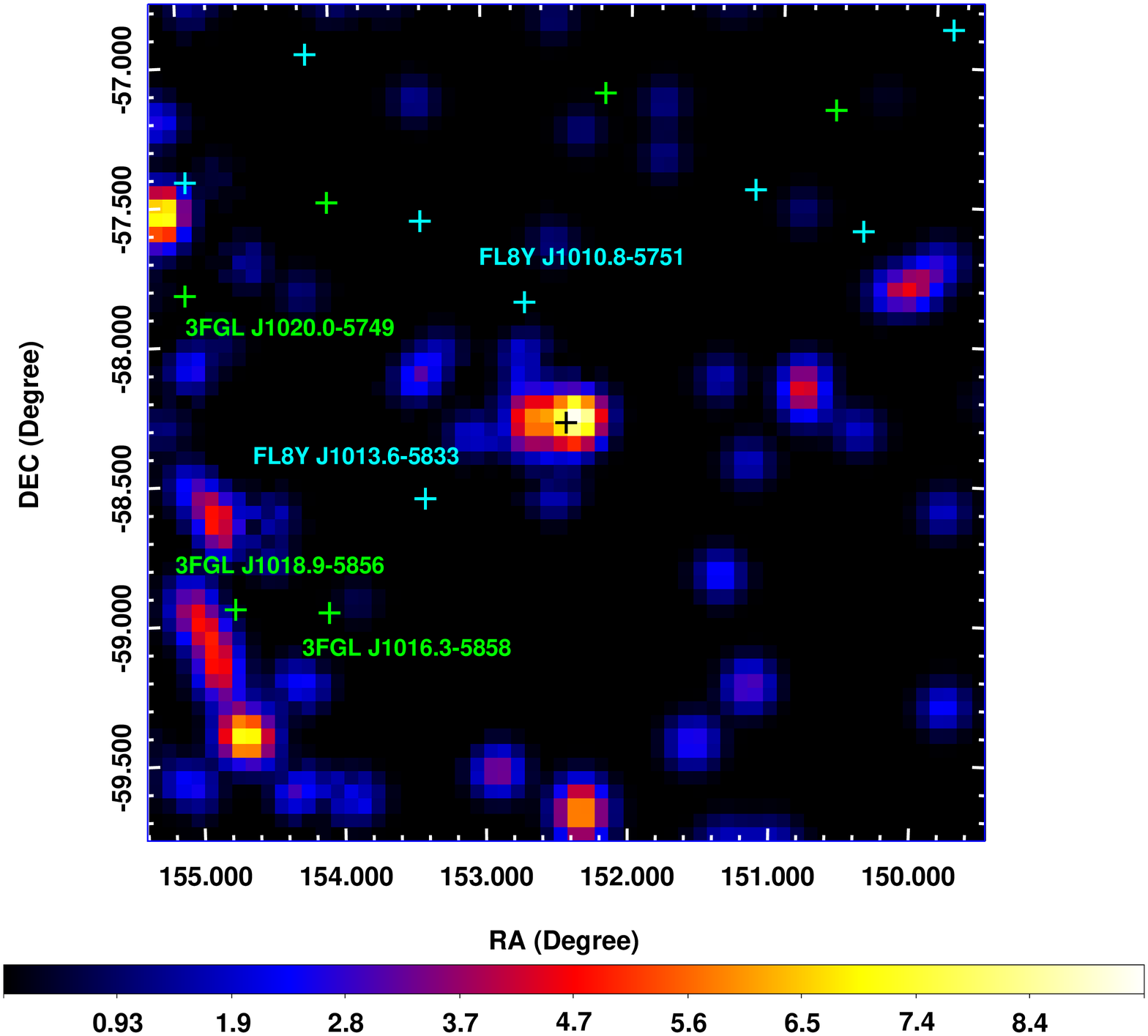}
\includegraphics[width=0.32\textwidth]{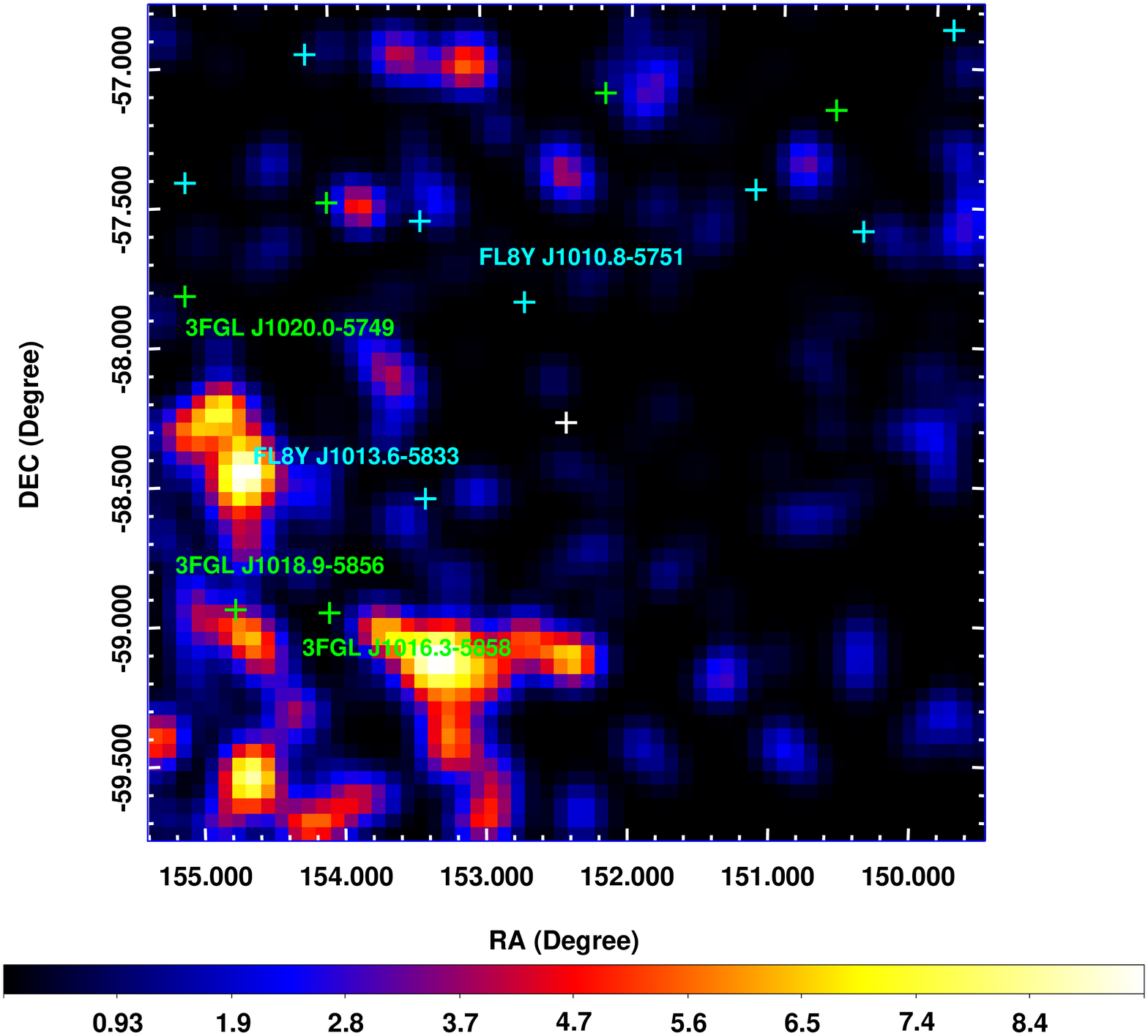}
\caption{
Residual TS maps for \gro, with a size of $\mathrm{3^{o}\times3^{o}}$, 
in 16.6--38.8 GeV band during Phase I, II, and the remaining phase ranges 
({\it left}, {\it middle}, and {\it right}, respectively). 
All sources in the source model (from the 4-year and 8-year Fermi LAT 
source catalogs) were considered and removed (marked with green and 
light blue pluses, respectively).
The image scale of the maps is 0\fdg05 pixel$^{-1}$, with the color bar 
indicating the TS value range. 
The position of \gro\ is in the center of the maps, marked with green
cross in the {\it left} panel and black or white plus in the other two panels.}
\label{fig:ts2}
\end{figure*}

We calculated the residual TS map in 16.6--38.8 GeV band during Phase II,
which is shown in the middle panel of Figure~\ref{fig:ts2}. 
The TS maps in this energy band
during Phase I and the remaining phase ranges were also
calculated, which are shown in the left and right panel
of Figure~\ref{fig:ts2} respectively. While the residual at the position
of \gro\ in Phase II has TS$\sim$10, the map is noisy with other residuals.
The situation is worse in Phase I, as a residual with TS$\sim$9 is slightly
off the position of \gro\ and there are several other residuals with similar
TS values. In the remaining phase ranges, TS$\sim$0 at the source position.
Based on the TS maps, we concluded that the result of the detection 
of \gro\ in Phase II is uncertain.

\begin{figure}
\centering
\epsscale{1.0}
\plotone{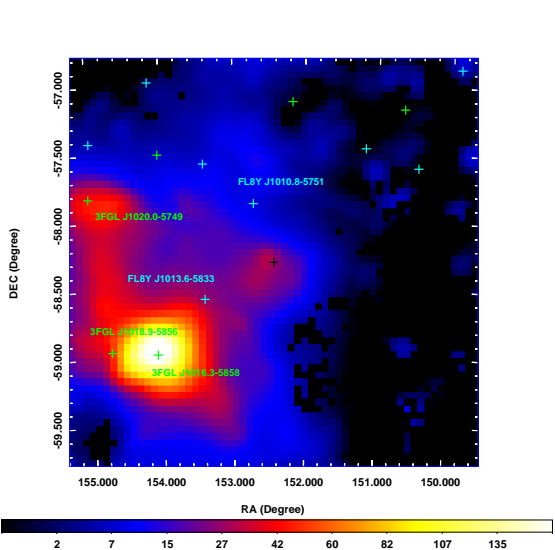}
\caption{Residual TS map for \gro\ in 0.5--39 GeV band during Phase I with 
the two nearby pulsars 3FGL J1016.3$-$5858 and J1020.0$-$5749 kept.
All other sources in the source model (from the 4-year and 8-year Fermi LAT 
source catalogs) were considered and removed (marked with green and light blue 
pluses, respectively). Excess emission at the position of \gro\ (marked with 
a black plus) is clearly resolved from the nearby pulsars.
The image scale of the map is 0\fdg05 pixel$^{-1}$, 
with the color bar indicating the TS value range.}
\label{fig:tsmap-2psr-phase1}
\end{figure}

\subsubsection{Possible Emission Contamination}                                                                                                                
\gro\ is located in a complex region with a few sources detected relatively
nearby and the $\gamma$-ray emission could only be detected during some 
phase ranges. We thus checked whether the nearby catalog sources are variable,
whose flux variations might affect our results.  There are six nearby sources 
listed in the LAT 4-year catalog (see Figure~\ref{fig:tsmap-fullband}). 
The brightest one (3FGL J1018.9$-$5856) is a \gr\ binary, two of them 
(3FGL J1016.3$-$5858 and 3FGL J1020.0$-$5749) are \gr\ pulsars 
(PSR J1016$-$5857 and PSR J1019$-$5749, respectively), and the other three 
are unidentified sources. We first checked the 0.1--300 GeV variability 
indices (\citealt{3fgl15}) of them, and found none of them has variability 
index greater than the threshold to be identified as a variable source.
There are also eight new sources listed in the LAT 8-year point source 
catalog (see Figure~\ref{fig:tsmap-fullband}). For them,
we calculated the 0.3--500 GeV variability indices for two of them 
(FL8Y J1010.8$-$5751 and FL8Y J1013.6$-$5833, the two closest to \gro) 
following the method used in \citet{3fgl15}. Both of these two sources 
are unidentified sources, and have variability indices 
(80.3 for J1010.8$-$5751 and 82.4 for J1013.6$-$5833) smaller than 
the threshold (149.7 for 112 degrees of freedom) considered for a variable 
source.
We then checked the 0.3--500 GeV fluxes of these nearby sources during 
the 10 orbital phase bins of \gro. For each of the sources, the fluxes 
during each bins have $<$2$\sigma$ deviations from those derived from 
the likelihood analysis of the whole data, again indicating
no significant variations.
We also checked the SIMBAD Astronomical Database and found that 
except \gro, there are only a few normal stars known within 
the 2$\sigma$ error circles of the position obtained during Phase I.
Therefore no evidence was found to indicate any possible emission 
contamination from nearby sources.

We also checked possible contamination due to the two nearby pulsars.
Ideally, we could gate off their pulsed emission for
this analysis. However no ephemerides covering the whole $>$9 years of
LAT observations are available for the two pulsars. We instead extracted 
the 0.5--39 GeV TS map during Phase I with these two pulsars kept
in the map.  The TS map (Figure~\ref{fig:tsmap-2psr-phase1}) shows that
excess \gr\ emission at the position of \gro\ is present with 
TS$\sim 28$ (comparing to TS$\sim 22$ when the two pulsars were removed; cf.
Figure~\ref{fig:tsmap-fullband}), and it is clearly resolved from that of 
the two pulsars; in other words, the excess is not due to non-clean removal of 
the pulsars. 

In addition, we evaluated how much the uncertainty of the Galactic
diffuse emission model, the dominant one among the systematic uncertainties, 
affected our detection result. By increasing the normalization of the model
by 6\%, we extracted the 0.5--39 GeV TS map during Phase I 
and found that the excess emission at the position of \gro\  is still present 
but with TS reduced to $\sim$18. The TS value corresponds to $\sim$4$\sigma$ 
detection significance, indicating that the detection result was not changed
by considering the systematic uncertainty.

\section{Discussion}
\label{sec:disc}

Having analyzed nine years of the Fermi LAT Pass 8 data, we searched for 
\gr\ emission from Be X-ray binaries with known orbital parameters.
The search was conducted in the whole data and orbital-phase resolved data.
Only for \gro, possible residual emission was found in the whole data, 
but without sufficiently high significance.
In the orbital-phase resolved analysis,
the \gro\ region was found to have excess \gr\ 
emission significantly detected during the orbital phase range of 0.8--0.9.
The \gr\ emission may be described with
a power law with $\Gamma\sim 2$, although the spectra suffer large 
uncertainties. The likely orbital dependence of the excess \gr\ emission 
supports its association with \gro. 
(In addition, given that the Fermi LAT collaboration very recently 
published an updated source catalog (4FGL catalog) and updated 
the diffuse background models, we checked whether our results would be
affected when the new source catalog and diffuse models were used.
Nearly the same results were obtained. We conclude that the results 
presented here are valid without significantly changes.)

%%No mechanisms or related physical processes are known to be able to produce \gr\ emission during neutron star's active accretion. The likely detection of \gro\  in phase 0.8--0.9, not during the periastron passage, is consistent with this fact. 
Since \gro\ is considered as a transient X-ray pulsar, with
pulsed X-ray emission detected during the outbursts,
accretion onto the neutron star certainly occurs at least in outbursts. 
When in quiescence, the X-ray luminosity of \gro\ is 
generally $>$10$^{34}$ erg s$^{-1}$, and in one {\it Swift} monitoring
of the binary over a full orbital cycle, the luminosity stayed at a level of
$\sim 10^{35}$ erg s$^{-1}$, higher than that for the onset of the propeller
phase \citep{tsy+17}. It is thus likely that there is always sufficiently
strong mass accretion in the accretion disk surrounding the neutron star, 
not allowing a possible 
switch for the neutron star from being accretion powered to rotation powered
(such as in the accretion-powered millisecond pulsar binary SAX J1808.4$-$3658;
for searches for \gr\ emission see \citealt{xw13,de+16}). 
The scenario for the
\gr\ production from PSR B1259$-$63/LS 2883 probably does not work for \gro.
\begin{table}
\centering
\tabletypesize{\footnotesize}
\tablecolumns{10}
\tablewidth{240pt}
\setlength{\tabcolsep}{2pt}
\caption{Three TS$>$9 detections in 0.1 orbital phase bins}
\begin{tabular}{lcc}
\hline
Central Time & Orbital Phase & TS \\
(MJD) & &  \\ \hline
55135.728 & 0.8--0.9 & 9.7 \\
55559.844 & 0.5--0.6 & 9.1 \\
56383.128 & 0.8--0.9  & 17.5 \\
\hline
\end{tabular}
\vskip 1mm
\label{tab:lc}
\end{table}

\begin{figure}
\centering
\epsscale{1.0}
\plotone{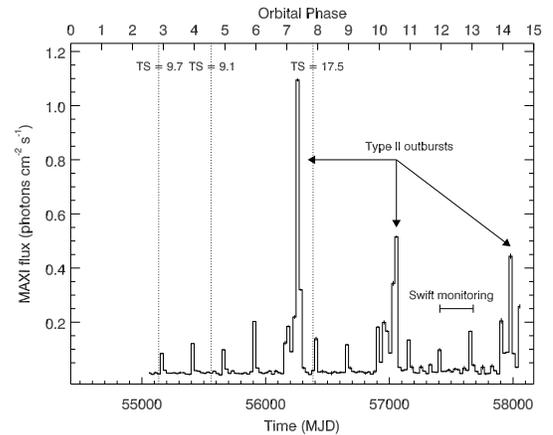}
\caption{\fermi\ LAT detections of \gro\ in three 0.1 orbital phases.
The {\it MAXI} \citep{maxi09} X-ray light curve of \gro\ is plotted to show
the outburst events. The occurrences of the three detections are marked
with dotted lines. The full orbital cycle observed with {\it Swift}
\citep{tsy+17} is also marked. }
\label{fig:lc}
\end{figure}

%These systems switch between the active state and disk-free state, and in the latter they are more like a typical MSP binary and do not have an accretion disk.  %%These features indicate that very complicated processes can occur in a seemingly accreting pulsar.

In addition to neutron star \gr\ binaries, another small group of so-called 
transitional millisecond pulsar (MSP) binary 
systems, with PSR J1023+0038 as a prototype \citep{arc+09}, are also known
to produce much enhanced \gr\ emission during their active state when 
an accretion disk is present around the pulsar \citep{wan+09,sta+14,tak+14}. 
Although the radiation mechanism for the enhanced $\gamma$-ray emission in
the active state is not certain, the scenarios of an active (or partially 
active; see 
\citealt{xwt18}) radio pulsar interacting with the accretion disk \citep{tak+14}
or a propellering neutron star \citep{pap+14,pt15} have been proposed.
Possible evidence for the former scenario is the coherent X-ray and optical pulsations detected in emission from PSR J1023+0038 in the active state (\citealt{arc+15,amb+17}; note no radio pulsations are seen).
However, the distance to \gro\ was estimated to be 
5.8$\pm$0.5\,kpc (\citealt{rtn12}; note its $V\simeq 15$\,mag, 
but no distance information is provided in Gaia Archive), based on which 
the \gr\ luminosity due to the detection in Phase~I
is $L_{\gamma}\simeq 3.6\times 10^{34}$\,erg\,s$^{-1}$ in 0.5--39 GeV (or 
$\simeq 6.7\times 10^{34}$\,erg\,s$^{-1}$ in 0.1--300 GeV). This high luminosity
value cannot be explained with the scenarios. Even assuming a pulsar wind
might be turned on in \gro, the long-term spin-down rate of the neutron star
would only provide a rotational power of 
$\sim 10^{31}$ erg\,s$^{-1}$ \citep{kuh+13}, too low to produce 
the observed \gr\ emission (J. Takata, private communication). 
Similarly in the propellering neutron star scenario, the predicted
\gr\ luminosity of \gro\ would be $\sim 10^{30}$ erg\,s$^{-1}$ 
(cf. Section 3.1 in \citealt{pt15}).

\begin{figure}
\centering
\epsscale{1.0}
\plotone{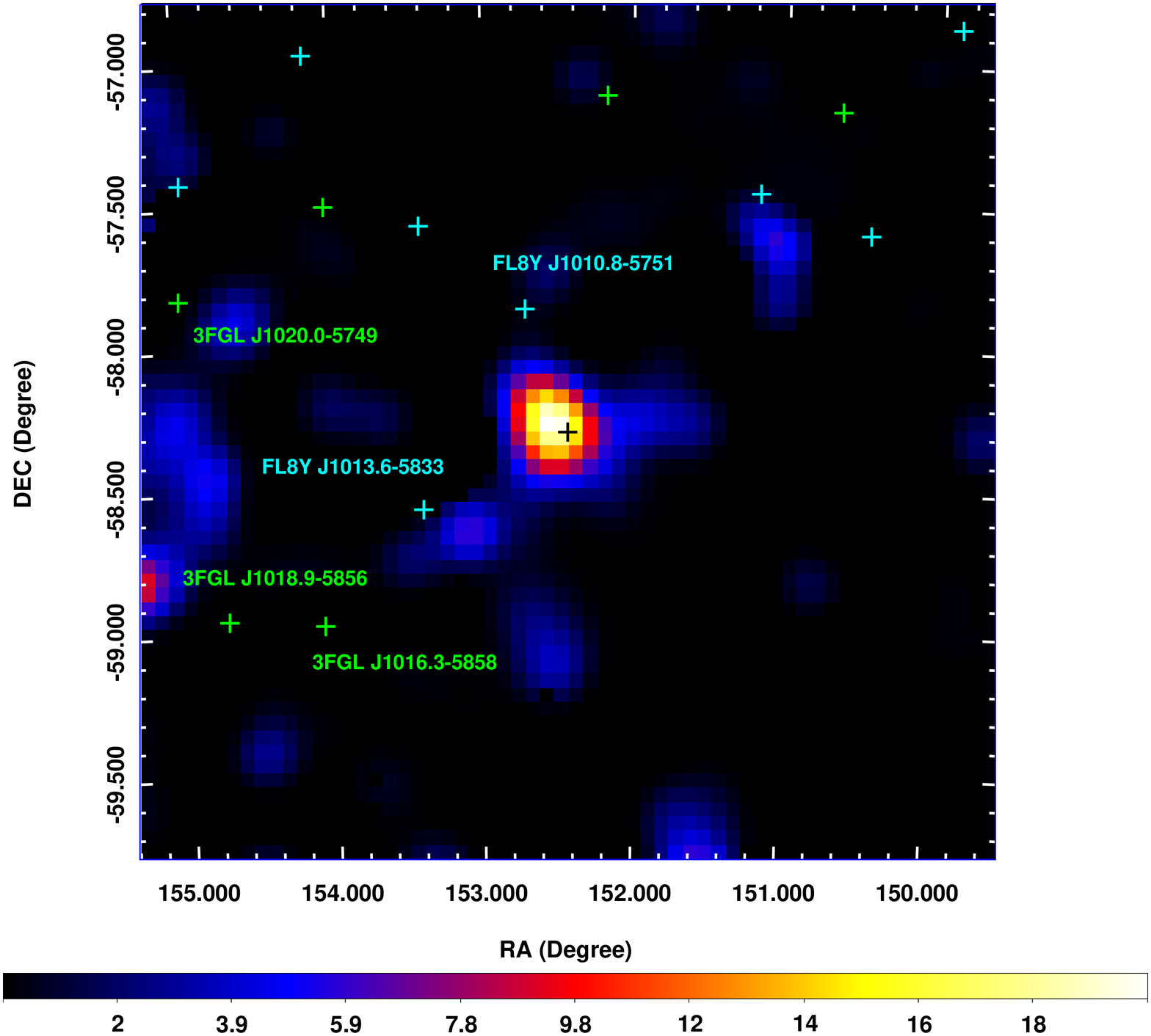}
\plotone{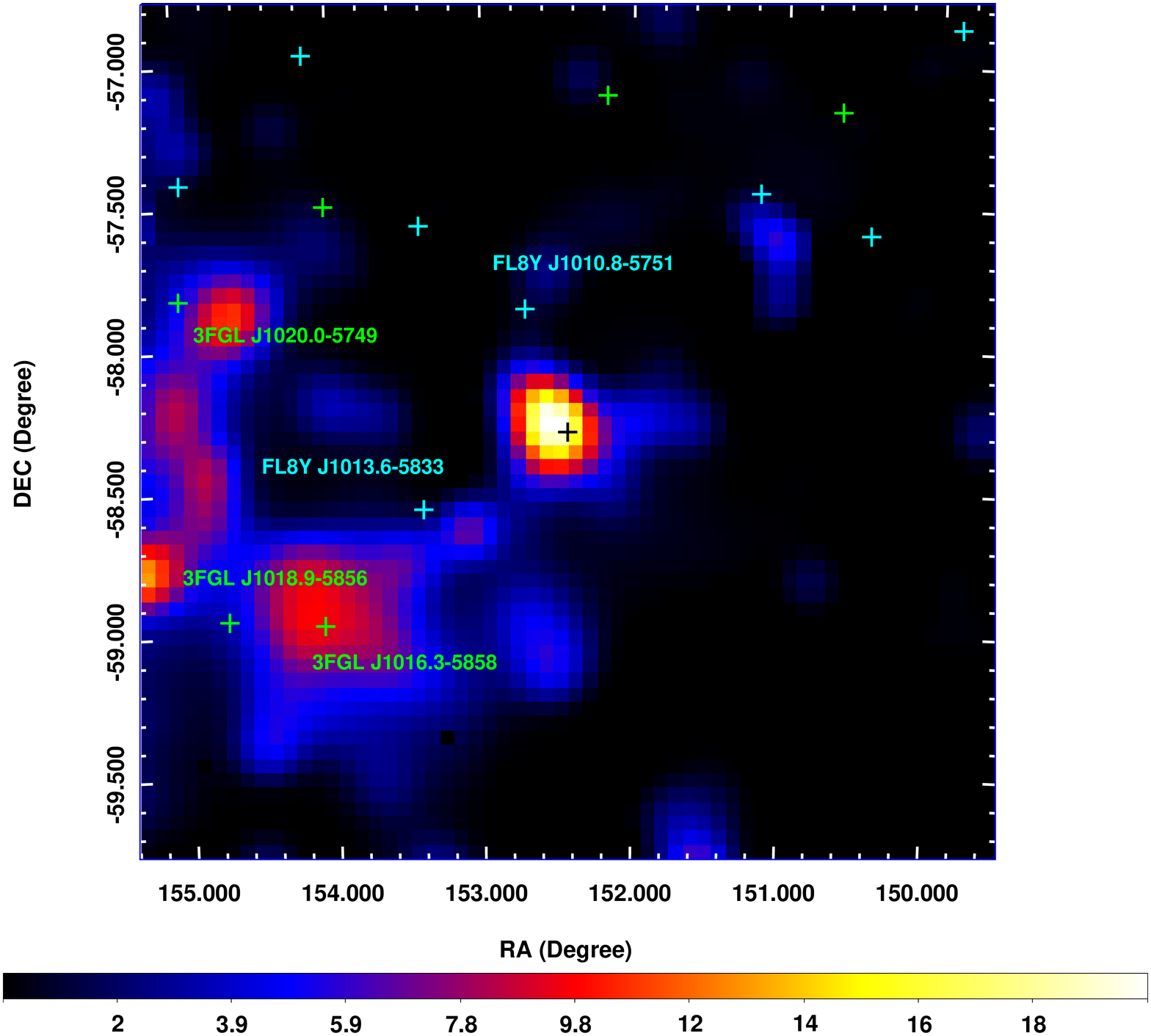}
\caption{TS maps of the \gro\ region during the single orbital phase 0.8--0.9
right after the 2012 November type-II outburst, with the two nearby 
pulsars 3FGL J1016.3$-$5858 and J1020.0$-$5749 removed and kept in the 
{\it upper} and {\it bottom} panels, respectively. Excess emission 
with TS=17.5 is clearly
seen at the position of \gro\ (marked with a black plus). 
All other sources in the source model (from the 4-year and 8-year Fermi LAT source 
catalogs) were considered and removed (marked with green and light blue 
pluses, respectively). The image scale of the maps is 0\fdg05 pixel$^{-1}$, 
with the color bars indicating the TS value range. }
\label{fig:tsd}
\end{figure}
%%The corotation radius $r_c$ of PSR J1023+0038 is small, $r_c\simeq 1.7\times 10^8 M_{1.4}^{1/3} P^{2/3}\ {\rm cm}\simeq 2.4\times 10^6$ cm, where $M_{1.4}$ is the neutron star mass in units of 1.4 $M_{\odot}$, and its magntospheric radius $r_m$ can be estimated to be of the order of 10$^7$ cm \citep{pt15}. On the basis of the accretion scenario for neutron stars generally considered \citep{fkr02}, PSR J1023+0038 is in the propeller phase in the active state, as $r_m > r_c$ (the inner edge of the disk does not reach the corotation radius).  For \gro, its $r_c\simeq 3.5\times 10^9$\,cm and $r_m\simeq 5.7\times 10^8 M_{1.4}^{1/7} R_6^{-2/7} L_{35}^{-2/7} \mu_{30}^{4/7} \simeq 1.9\times 10^9 L_{35}^{-2/7}$\,cm, where the neutron star radius $R_6$ is in units of $10^6$\,cm, $L_{35}$ is the X-ray luminosity in units of 10$^{35}$\,erg\,s$^{-1}$, and $\mu_{30}$ is the magnetic moment in units of 10$^{30}$\,G\,cm$^3$ ($\mu_{30}=8$ assumed for \gro).  Because $r_m$ is slightly smaller than $r_c$, \gro\ would switch to be in the propeller phase once its $L$ is lower than 10$^{34}$\,erg\,s$^{-1}$ (see also \citealt{tsy+17}), opening the possibility to be in the \gr\ emitting scenarios suggested by \citet{pt15} or \citet{tak+14}.

Among previous searches for \gr\ emission from neutron star X-ray binaries
(e.g., \citealt{xw13,de+16}),
\citet{li+12} have provided a very similar case to \gro\ by justifying that
the \gr\ transients GRO J1036$-$55 and AGL J1037$-$5708 are associated with
the HMXB 4U~1036$-$56. The likely association indicates that the HMXB 
only had occasional (detectable) \gr\ emission during several days. 
Following this feature, we performed the likelihood analysis
to the LAT data in each time intervals of 0.1 orbital phase of \gro\,
(i.e., 24.9 days). We found that there were three possible 
detection events with TS$>$9, which are
listed in Table~\ref{tab:lc}. Two were in phase 0.8--0.9 and
one in 0.5--0.6, and their occurrences are shown in Figure~\ref{fig:lc}, 
comparing to the X-ray outburst activity. It can be seen that the detection
with the highest TS value of 17.5 was in the phase of 0.8--0.9, 
after a bright type-II outburst (occurred in November 2012 at an orbital
phase of $\sim$0.3) during the same orbital cycle. The other two
possible events only had TS$\simeq$9, too marginal to be considered.
We made the TS
maps for the TS=17.5 detection and showed them in Figure~\ref{fig:tsd},
indicating clear excess emission at the position of \gro.
A fine light curve (5 day binned) during this orbital phase range of
0.8--0.9 was constructed, 
but no clear pattern, such as a flaring event, could be identified (which
actually excludes possibilities of contamination by a Solar flare or a
$\gamma$-ray burst; both would appear as bright and sharp flaring events 
in daily LAT light curves). Therefore similar to that seen in 
4U~1036$-$56, 
the occurences of the \gr\ emitting events did not coincide with any X-ray 
outbursts.

In \citet{bed09}, a model for \gr\ emission from neutron star HMXBs was 
proposed, in which a turbulent and magnetized region near an accreting 
neutron star would be a site accelerating particles to relativistic 
energies. \citet{li+12} applied this model to 4U 1036-56, but the energy, 
considered to be transferred from the rotating neutron star to the high-energy 
particles in the model, is not large enough for the observed \gr\ emission 
from the HMXB, and thus they suggested that the available 
accretion energy should be considered. For \gro, the same situation presents.
The available power $L_{\rm rot}$ from the rotating neutron star is low,
$L_{\rm rot}\sim 3\times 10^{30} B^{8/7}_{12}\dot{M}^{3/7}_{16}P^{-2}_{100}$\,erg\,s$^{-1}$ \citep{bed09,li+12},
where $B_{12}$ is the magnetic field of the neutron star in units of 
$10^{12}$\,G, $\dot{M}_{16}$ is the mass accretion rate in units of 
$10^{16}$\,g\,s$^{-1}$, and $P_{100}$ is the spin period in units of 100 sec.
The accretion power at the magnetospheric radius (i.e., at which
the accretion disk is truncated by the magnetic field of the neutron star) is 
$L_{\rm acc}\sim 4.6\times 10^{33}B^{-4/7}_{12}\dot{M}^{9/7}_{16}$\,erg\,s$^{-1}$ \citep{li+12};
when $\dot{M}_{16} \gtrsim 12$, $L_{\rm acc} \gtrsim L_{\gamma}$.
Note that both 4U 1036$-$56 and \gro\ require high mass accretion rates
in order to match the observed \gr\ luminosities, if the face values
of the estimated source distances are used.

We conclude that the \gr\ emission likely from \gro\ was detected with the \fermi\ LAT. 
Similar to that in 4U 1036$-$56, the emission was not persistent but 
occasionally detectable. The detailed physical process that gives rise to
such high-energy emission is not clear, as the current models considering an 
accreting neutron star have difficulty in explaining the observed relatively
high luminosities. Further observational studies of these HMXBs at high-energies
should help establish their emission properties more clearly and may provide
a clue for identifying the emission mechanism.

\acknowledgements
This research made use of the High Performance Computing Resource in the Core
Facility for Advanced Research Computing at Shanghai Astronomical Observatory,
and also made use of the MAXI data provided by RIKEN, JAXA and the MAXI team.
This research was supported by the National Program on Key Research 
and Development Project (Grant No. 2016YFA0400804) and
the National Natural Science Foundation
of China (U1738131, 11633007).


\begin{thebibliography}{48}
\expandafter\ifx\csname natexlab\endcsname\relax\def\natexlab#1{#1}\fi

\bibitem[{{Abdo} {et~al.}(2009){Abdo}, {Ackermann}, {Ajello}, {Atwood},
  {Axelsson}, {Baldini}, {Ballet}, {Barbiellini}, {Bastieri}, {Baughman},
  {Bechtol}, {Bellazzini}, {Berenji}, {Blandford}, {Bloom}, {Bonamente},
  {Borgland}, {Bregeon}, {Brez}, {Brigida}, {Bruel}, {Burnett}, {Caliandro},
  {Cameron}, {Caraveo}, {Casandjian}, {Cavazzuti}, {Cecchi}, {{\c C}elik},
  {Charles}, {Chaty}, {Chekhtman}, {Cheung}, {Chiang}, {Ciprini}, {Claus},
  {Cohen-Tanugi}, {Cominsky}, {Conrad}, {Corbel}, {Corbet}, {Cutini}, {Dermer},
  {de Angelis}, {de Luca}, {de Palma}, {Digel}, {Dormody}, {do Couto e Silva},
  {Drell}, {Dubois}, {Dubus}, {Dumora}, {Farnier}, {Favuzzi}, {Fegan}, {Focke},
  {Frailis}, {Fukazawa}, {Funk}, {Fusco}, {Gargano}, {Gasparrini}, {Gehrels},
  {Germani}, {Giebels}, {Giglietto}, {Giordano}, {Glanzman}, {Godfrey},
  {Grenier}, {Grondin}, {Grove}, {Guillemot}, {Guiriec}, {Hanabata}, {Harding},
  {Hayashida}, {Hays}, {Hill}, {Hughes}, {J{\'o}hannesson}, {Johnson},
  {Johnson}, {Johnson}, {Johnson}, {Kamae}, {Katagiri}, {Kataoka}, {Kawai},
  {Kerr}, {Kn{\"o}dlseder}, {Kocian}, {Kuehn}, {Kuss}, {Lande}, {Larsson},
  {Latronico}, {Longo}, {Loparco}, {Lott}, {Lovellette}, {Lubrano}, {Madejski},
  {Makeev}, {Marelli}, {Mazziotta}, {McEnery}, {Meurer}, {Michelson},
  {Mitthumsiri}, {Mizuno}, {Monte}, {Monzani}, {Morselli}, {Moskalenko},
  {Murgia}, {Nolan}, {Nuss}, {Ohsugi}, {Okumura}, {Omodei}, {Orlando}, {Ormes},
  {Paneque}, {Panetta}, {Parent}, {Pelassa}, {Pepe}, {Pesce-Rollins}, {Piron},
  {Porter}, {Rain{\`o}}, {Rando}, {Ray}, {Razzano}, {Rea}, {Reimer}, {Reimer},
  {Reposeur}, {Ritz}, {Rochester}, {Rodriguez}, {Romani}, {Ryde},
  {Sadrozinski}, {Sanchez}, {Sander}, {Saz Parkinson}, {Scargle}, {Sgr{\`o}},
  {Shaw}, {Sierpowska-Bartosik}, {Siskind}, {Smith}, {Smith}, {Spandre},
  {Spinelli}, {Striani}, {Strickman}, {Suson}, {Tajima}, {Takahashi},
  {Takahashi}, {Tanaka}, {Thayer}, {Thayer}, {Thompson}, {Tibaldo}, {Torres},
  {Tosti}, {Tramacere}, {Uchiyama}, {Usher}, {Vasileiou}, {Vilchez}, {Vitale},
  {Waite}, {Wang}, {Winer}, {Wood}, {Ylinen}, \& {Ziegler}}]{abd+09}
{Abdo}, A.~A., {Ackermann}, M., {Ajello}, M., {et~al.} 2009, \apjl, 701, L123

\bibitem[{{Abdo} {et~al.}(2010){Abdo}, {Ackermann}, {Ajello}, {Allafort},
  {Antolini}, {Atwood}, {Axelsson}, {Baldini}, {Ballet}, {Barbiellini}, \&
  et~al.}]{1fgl}
---. 2010, \apjs, 188, 405

\bibitem[{{Abdo} {et~al.}(2011){Abdo}, {Ackermann}, {Ajello}, {Allafort},
  {Ballet}, {Barbiellini}, {Bastieri}, {Bechtol}, {Bellazzini}, {Berenji},
  {Blandford}, {Bonamente}, {Borgland}, {Bregeon}, {Brigida}, {Bruel},
  {Buehler}, {Buson}, {Caliandro}, {Cameron}, {Camilo}, {Caraveo}, {Cecchi},
  {Charles}, {Chaty}, {Chekhtman}, {Chernyakova}, {Cheung}, {Chiang},
  {Ciprini}, {Claus}, {Cohen-Tanugi}, {Cominsky}, {Corbel}, {Cutini},
  {D'Ammando}, {de Angelis}, {den Hartog}, {de Palma}, {Dermer}, {Digel},
  {Silva}, {Dormody}, {Drell}, {Drlica-Wagner}, {Dubois}, {Dubus}, {Dumora},
  {Enoto}, {Espinoza}, {Favuzzi}, {Fegan}, {Ferrara}, {Focke}, {Fortin},
  {Fukazawa}, {Funk}, {Fusco}, {Gargano}, {Gasparrini}, {Gehrels}, {Germani},
  {Giglietto}, {Giommi}, {Giordano}, {Giroletti}, {Glanzman}, {Godfrey},
  {Grenier}, {Grondin}, {Grove}, {Grundstrom}, {Guiriec}, {Gwon}, {Hadasch},
  {Harding}, {Hayashida}, {Hays}, {J{\'o}hannesson}, {Johnson}, {Johnson},
  {Johnston}, {Kamae}, {Katagiri}, {Kataoka}, {Keith}, {Kerr},
  {Kn{\"o}dlseder}, {Kramer}, {Kuss}, {Lande}, {Lee}, {Lemoine-Goumard},
  {Longo}, {Loparco}, {Lovellette}, {Lubrano}, {Manchester}, {Marelli},
  {Mazziotta}, {Michelson}, {Mitthumsiri}, {Mizuno}, {Moiseev}, {Monte},
  {Monzani}, {Morselli}, {Moskalenko}, {Murgia}, {Nakamori}, {Naumann-Godo},
  {Neronov}, {Nolan}, {Norris}, {Noutsos}, {Nuss}, {Ohsugi}, {Okumura},
  {Omodei}, {Orlando}, {Paneque}, {Parent}, {Pesce-Rollins}, {Pierbattista},
  {Piron}, {Porter}, {Possenti}, {Rain{\`o}}, {Rando}, {Ray}, {Razzano},
  {Razzaque}, {Reimer}, {Reimer}, {Reposeur}, {Ritz}, {Sadrozinski}, {Scargle},
  {Sgr{\`o}}, {Shannon}, {Siskind}, {Smith}, {Spandre}, {Spinelli},
  {Strickman}, {Suson}, {Takahashi}, {Tanaka}, {Thayer}, {Thayer}, {Thompson},
  {Thorsett}, {Tibaldo}, {Tibolla}, {Torres}, {Tosti}, {Troja}, {Uchiyama},
  {Usher}, {Vandenbroucke}, {Vasileiou}, {Vianello}, {Vitale}, {Waite}, {Wang},
  {Winer}, {Wolff}, {Wood}, {Wood}, {Yang}, {Ziegler}, \& {Zimmer}}]{abd+11}
---. 2011, \apjl, 736, L11

\bibitem[{{Acciari} {et~al.}(2011){Acciari}, {Aliu}, {Araya}, {Arlen}, {Aune},
  {Beilicke}, {Benbow}, {Bradbury}, {Buckley}, {Bugaev}, {Byrum}, {Cannon},
  {Cesarini}, {Ciupik}, {Collins-Hughes}, {Cui}, {Dickherber}, {Duke},
  {Falcone}, {Finley}, {Fortson}, {Furniss}, {Galante}, {Gall}, {Godambe},
  {Griffin}, {Guenette}, {Gyuk}, {Hanna}, {Holder}, {Hughes}, {Hui},
  {Humensky}, {Imran}, {Kaaret}, {Kertzman}, {Krawczynski}, {Krennrich},
  {Madhavan}, {Maier}, {Majumdar}, {McArthur}, {Moriarty}, {Ong}, {Otte},
  {Pandel}, {Park}, {Perkins}, {Pohl}, {Prokoph}, {Quinn}, {Ragan}, {Reyes},
  {Reynolds}, {Roache}, {Rose}, {Saxon}, {Sembroski}, {{\c S}ent{\"u}rk},
  {Smith}, {Te{\v s}i{\'c}}, {Theiling}, {Thibadeau}, {Varlotta}, {Vincent},
  {Vivier}, {Wakely}, {Ward}, {Weekes}, {Weinstein}, {Weisgarber}, {Weng},
  {Williams}, {Wood}, \& {Zitzer}}]{acc+11}
{Acciari}, V.~A., {Aliu}, E., {Araya}, M., {et~al.} 2011, \apj, 733, 96

\bibitem[{{Acero} {et~al.}(2015){Acero}, {Ackermann}, {Ajello}, {Albert},
  {Atwood}, {Axelsson}, {Baldini}, {Ballet}, {Barbiellini}, {Bastieri},
  {Belfiore}, {Bellazzini}, {Bissaldi}, {Blandford}, {Bloom}, {Bogart},
  {Bonino}, {Bottacini}, {Bregeon}, {Britto}, {Bruel}, {Buehler}, {Burnett},
  {Buson}, {Caliandro}, {Cameron}, {Caputo}, {Caragiulo}, {Caraveo},
  {Casandjian}, {Cavazzuti}, {Charles}, {Chaves}, {Chekhtman}, {Cheung},
  {Chiang}, {Chiaro}, {Ciprini}, {Claus}, {Cohen-Tanugi}, {Cominsky}, {Conrad},
  {Cutini}, {D'Ammando}, {de Angelis}, {DeKlotz}, {de Palma}, {Desiante},
  {Digel}, {Di Venere}, {Drell}, {Dubois}, {Dumora}, {Favuzzi}, {Fegan},
  {Ferrara}, {Finke}, {Franckowiak}, {Fukazawa}, {Funk}, {Fusco}, {Gargano},
  {Gasparrini}, {Giebels}, {Giglietto}, {Giommi}, {Giordano}, {Giroletti},
  {Glanzman}, {Godfrey}, {Grenier}, {Grondin}, {Grove}, {Guillemot}, {Guiriec},
  {Hadasch}, {Harding}, {Hays}, {Hewitt}, {Hill}, {Horan}, {Iafrate}, {Jogler},
  {J{\'o}hannesson}, {Johnson}, {Johnson}, {Johnson}, {Johnson}, {Kamae},
  {Kataoka}, {Katsuta}, {Kuss}, {La Mura}, {Landriu}, {Larsson}, {Latronico},
  {Lemoine-Goumard}, {Li}, {Li}, {Longo}, {Loparco}, {Lott}, {Lovellette},
  {Lubrano}, {Madejski}, {Massaro}, {Mayer}, {Mazziotta}, {McEnery},
  {Michelson}, {Mirabal}, {Mizuno}, {Moiseev}, {Mongelli}, {Monzani},
  {Morselli}, {Moskalenko}, {Murgia}, {Nuss}, {Ohno}, {Ohsugi}, {Omodei},
  {Orienti}, {Orlando}, {Ormes}, {Paneque}, {Panetta}, {Perkins},
  {Pesce-Rollins}, {Piron}, {Pivato}, {Porter}, {Racusin}, {Rando}, {Razzano},
  {Razzaque}, {Reimer}, {Reimer}, {Reposeur}, {Rochester}, {Romani},
  {Salvetti}, {S{\'a}nchez-Conde}, {Saz Parkinson}, {Schulz}, {Siskind},
  {Smith}, {Spada}, {Spandre}, {Spinelli}, {Stephens}, {Strong}, {Suson},
  {Takahashi}, {Takahashi}, {Tanaka}, {Thayer}, {Thayer}, {Thompson},
  {Tibaldo}, {Tibolla}, {Torres}, {Torresi}, {Tosti}, {Troja}, {Van Klaveren},
  {Vianello}, {Winer}, {Wood}, {Wood}, {Zimmer}, \& {Fermi-LAT
  Collaboration}}]{3fgl15}
{Acero}, F., {Ackermann}, M., {Ajello}, M., {et~al.} 2015, \apjs, 218, 23

\bibitem[Ackermann et al.(2013)]{ack+13} Ackermann, M., Ajello, M., Ballet, J., et al.\ 2013, \apjl, 773, L35 

\bibitem[{{Aharonian} {et~al.}(2009){Aharonian}, {Akhperjanian}, {Anton},
  {Barres de Almeida}, {Bazer-Bachi}, {Becherini}, {Behera}, {Bernl{\"o}hr},
  {Bochow}, {Boisson}, {Bolmont}, {Borrel}, {Brucker}, {Brun}, {Brun},
  {B{\"u}hler}, {Bulik}, {B{\"u}sching}, {Boutelier}, {Chadwick},
  {Charbonnier}, {Chaves}, {Cheesebrough}, {Chounet}, {Clapson}, {Coignet},
  {Dalton}, {Daniel}, {Davids}, {Degrange}, {Deil}, {Dickinson},
  {Djannati-Ata{\"i}}, {Domainko}, {O'C.~Drury}, {Dubois}, {Dubus}, {Dyks},
  {Dyrda}, {Egberts}, {Emmanoulopoulos}, {Espigat}, {Farnier}, {Feinstein},
  {Fiasson}, {F{\"o}rster}, {Fontaine}, {F{\"u}{\ss}ling}, {Gabici}, {Gallant},
  {G{\'e}rard}, {Gerbig}, {Giebels}, {Glicenstein}, {Gl{\"u}ck}, {Goret},
  {G{\"o}ring}, {Hauser}, {Hauser}, {Heinz}, {Heinzelmann}, {Henri}, {Hermann},
  {Hinton}, {Hoffmann}, {Hofmann}, {Holleran}, {Hoppe}, {Horns},
  {Jacholkowska}, {de Jager}, {Jahn}, {Jung}, {Katarzy{\'n}ski}, {Katz},
  {Kaufmann}, {Kerschhaggl}, {Khangulyan}, {Kh{\'e}lifi}, {Keogh}, {Klochkov},
  {Klu{\'z}niak}, {Kneiske}, {Komin}, {Kosack}, {Kossakowski}, {Lamanna},
  {Lenain}, {Lohse}, {Marandon}, {Martineau-Huynh}, {Marcowith}, {Masbou},
  {Maurin}, {McComb}, {Medina}, {Moderski}, {Moulin}, {Naumann-Godo}, {de
  Naurois}, {Nedbal}, {Nekrassov}, {Nicholas}, {Niemiec}, {Nolan}, {Ohm},
  {Olive}, {de O{\~n}a Wilhelmi}, {Orford}, {Ostrowski}, {Panter}, {Paz
  Arribas}, {Pedaletti}, {Pelletier}, {Petrucci}, {Pita}, {P{\"u}hlhofer},
  {Punch}, {Quirrenbach}, {Raubenheimer}, {Raue}, {Rayner}, {Renaud}, {Rieger},
  {Ripken}, {Rob}, {Rosier-Lees}, {Rowell}, {Rudak}, {Rulten}, {Ruppel},
  {Sahakian}, {Santangelo}, {Schlickeiser}, {Sch{\"o}ck}, {Schwanke},
  {Schwarzburg}, {Schwemmer}, {Shalchi}, {Sikora}, {Skilton}, {Sol},
  {Spangler}, {Stawarz}, {Steenkamp}, {Stegmann}, {Stinzing}, {Superina},
  {Szostek}, {Tam}, {Tavernet}, {Terrier}, {Tibolla}, {Tluczykont}, {van
  Eldik}, {Vasileiadis}, {Venter}, {Venter}, {Vialle}, {Vincent}, {Vivier},
  {V{\"o}lk}, {Volpe}, {Wagner}, {Ward}, {Zdziarski}, \& {Zech}}]{aha+09}
{Aharonian}, F., {Akhperjanian}, A.~G., {Anton}, G., {et~al.} 2009, \aap, 507,
  389
  
\bibitem[Ahnen et al.(2016)]{ahn+16} Ahnen, M.~L., Ansoldi, S., Antonelli, L.~A., et al.\ 2016, \aap, 591, A76 

\bibitem[{{Ambrosino} {et~al.}(2017){Ambrosino}, {Papitto}, {Stella}, {Meddi},
  {Cretaro}, {Burderi}, {Di Salvo}, {Israel}, {Ghedina}, {Di Fabrizio}, \&
  {Riverol}}]{amb+17}
{Ambrosino}, F., {Papitto}, A., {Stella}, L., {et~al.} 2017, Nature Astronomy,
  1, 854
  
\bibitem[Anchordoqui et al.(2003)]{anc+03} Anchordoqui, L.~A., Torres, D.~F., McCauley, T.~P., Romero, G.~E., \& Aharonian, F.~A.\ 2003, \apj, 589, 481

\bibitem[{{Archibald} {et~al.}(2009){Archibald}, {Stairs}, {Ransom}, {Kaspi},
  {Kondratiev}, {Lorimer}, {McLaughlin}, {Boyles}, {Hessels}, {Lynch}, {van
  Leeuwen}, {Roberts}, {Jenet}, {Champion}, {Rosen}, {Barlow}, {Dunlap}, \&
  {Remillard}}]{arc+09}
{Archibald}, A.~M., {Stairs}, I.~H., {Ransom}, S.~M., {et~al.} 2009, Science,
  324, 1411

\bibitem[{{Archibald} {et~al.}(2015){Archibald}, {Bogdanov}, {Patruno},
  {Hessels}, {Deller}, {Bassa}, {Janssen}, {Kaspi}, {Lyne}, {Stappers},
  {Tendulkar}, {D'Angelo}, \& {Wijnands}}]{arc+15}
{Archibald}, A.~M., {Bogdanov}, S., {Patruno}, A., {et~al.} 2015, \apj, 807, 62

\bibitem[{{Atwood} {et~al.}(2009){Atwood}, {Abdo}, {Ackermann}, {Althouse},
  {Anderson}, {Axelsson}, {Baldini}, {Ballet}, {Band}, {Barbiellini}, \&
  et~al.}]{atw+09}
{Atwood}, W.~B., {Abdo}, A.~A., {Ackermann}, M., {et~al.} 2009, \apj, 697, 1071

\bibitem[{Bednarek}(2009)]{bed09}
{Bednarek}, W. 2009, \aap, 495, 919

\bibitem[{{Bongiorno} {et~al.}(2011){Bongiorno}, {Falcone}, {Stroh}, {Holder},
  {Skilton}, {Hinton}, {Gehrels}, \& {Grube}}]{bon+11}
{Bongiorno}, S.~D., {Falcone}, A.~D., {Stroh}, M., {et~al.} 2011, \apjl, 737,
  L11

\bibitem[{{Burderi} {et~al.}(2003){Burderi}, {Di Salvo}, {D'Antona}, {Testa},
  {Iaria}, {Lavagetto}, \& {Robba}}]{bur+03}
{Burderi}, L., {Di Salvo}, T., {D'Antona}, F., {et~al.} 2003, Chinese Journal
  of Astronomy and Astrophysics Supplement, 3, 311

\bibitem[{{Cheng} \& {Ruderman}(1991)}]{cr91}
{Cheng}, K.~S., \& {Ruderman}, M. 1991, \apj, 373, 187

\bibitem[{{Chernyakova} {et~al.}(2014){Chernyakova}, {Abdo}, {Neronov},
  {McSwain}, {Mold{\'o}n}, {Rib{\'o}}, {Paredes}, {Sushch}, {de Naurois},
  {Schwanke}, {Uchiyama}, {Wood}, {Johnston}, {Chaty}, {Coleiro}, {Malyshev},
  \& {Babyk}}]{che+14}
{Chernyakova}, M., {Abdo}, A.~A., {Neronov}, A., {et~al.} 2014, \mnras, 439,
  432

\bibitem[{{Coe} {et~al.}(2007){Coe}, {Bird}, {Hill}, {McBride}, {Schurch},
  {Galache}, {Wilson}, {Finger}, {Buckley}, \& {Romero-Colmenero}}]{coe+07}
{Coe}, M.~J., {Bird}, A.~J., {Hill}, A.~B., {et~al.} 2007, \mnras, 378, 1427

\bibitem[{{Davidson} \& {Ostriker}(1973)}]{do73}
{Davidson}, K., \& {Ostriker}, J.~P. 1973, \apj, 179, 585

\bibitem[{{de O{\~n}a Wilhelmi} {et~al.}(2016)}]{de+16}
{de O{\~n}a Wilhelmi}, E., {Papitto}, A., {Li}, J., et al. 2016, \mnras,
456, 2647

\bibitem[{{Dubus}(2013)}]{dub13}
{Dubus}, G. 2013, \aapr, 21, 64

\bibitem[{{Frank} {et~al.}(2002){Frank}, {King}, \& {Raine}}]{fkr02}
{Frank}, J., {King}, A., \& {Raine}, D.~J. 2002, {Accretion Power in
  Astrophysics: Third Edition}, 398

\bibitem[{{Hadasch} {et~al.}(2012){Hadasch}, {Torres}, {Tanaka}, {Corbet},
  {Hill}, {Dubois}, {Dubus}, {Glanzman}, {Corbel}, {Li}, {Chen}, {Zhang},
  {Caliandro}, {Kerr}, {Richards}, {Max-Moerbeck}, {Readhead}, \&
  {Pooley}}]{had2012}
{Hadasch}, D., {Torres}, D.~F., {Tanaka}, T., {et~al.} 2012, \apj, 749, 54

\bibitem[{{Hinton} {et~al.}(2009){Hinton}, {Skilton}, {Funk}, {Brucker},
  {Aharonian}, {Dubus}, {Fiasson}, {Gallant}, {Hofmann}, {Marcowith}, \&
  {Reimer}}]{hin+09}
{Hinton}, J.~A., {Skilton}, J.~L., {Funk}, S., {et~al.} 2009, \apjl, 690, L101

\bibitem[{{Ho} {et~al.}(2017){Ho}, {Ng}, {Lyne}, {Stappers}, {Coe}, {Halpern},
  {Johnson}, \& {Steele}}]{ho+17}
{Ho}, W.~C.~G., {Ng}, C.-Y., {Lyne}, A.~G., {et~al.} 2017, \mnras, 464, 1211

\bibitem[{{Johnston} {et~al.}(1992){Johnston}, {Manchester}, {Lyne}, {Bailes},
  {Kaspi}, {Qiao}, \& {D'Amico}}]{joh+92}
{Johnston}, S., {Manchester}, R.~N., {Lyne}, A.~G., {et~al.} 1992, \apjl, 387,
  L37

\bibitem[{{Johnston} {et~al.}(1994){Johnston}, {Manchester}, {Lyne},
  {Nicastro}, \& {Spyromilio}}]{joh+94}
{Johnston}, S., {Manchester}, R.~N., {Lyne}, A.~G., {Nicastro}, L., \&
  {Spyromilio}, J. 1994, \mnras, 268, 430

\bibitem[{{K{\"u}hnel} {et~al.}(2013){K{\"u}hnel}, {M{\"u}ller}, {Kreykenbohm},
  {F{\"u}rst}, {Pottschmidt}, {Rothschild}, {Caballero}, {Grinberg},
  {Sch{\"o}nherr}, {Shrader}, {Klochkov}, {Staubert}, {Ferrigno},
  {Torrej{\'o}n}, {Mart{\'{\i}}nez-N{\'u}{\~n}ez}, \& {Wilms}}]{kuh+13}
{K{\"u}hnel}, M., {M{\"u}ller}, S., {Kreykenbohm}, I., {et~al.} 2013, \aap,
  555, A95

\bibitem[{{K{\"u}hnel} {et~al.}(2017){K{\"u}hnel}, {F{\"u}rst}, {Pottschmidt},
  {Kreykenbohm}, {Ballhausen}, {Falkner}, {Rothschild}, {Klochkov}, \&
  {Wilms}}]{kuh+17}
{K{\"u}hnel}, M., {F{\"u}rst}, F., {Pottschmidt}, K., {et~al.} 2017, \aap, 607,
  A88

\bibitem[{{Lamers} {et~al.}(1976){Lamers}, {van den Heuvel}, \&
  {Petterson}}]{lvp76}
{Lamers}, H.~J.~G.~L.~M., {van den Heuvel}, E.~P.~J., \& {Petterson}, J.~A.
  1976, \aap, 49, 327

\bibitem[{{Li} {et~al.}(2012)}]{li+12}
{Li}, J., {Torres}, D.~F., {Zhang}, S., {Papitto}, A., {Chen}, Y., 
{Wang}, J.-M. 2012, \apj, 761, 49

\bibitem[{{Li} {et~al.}(2017){Li}, {Torres}, {Cheng}, {de O{\~n}a Wilhelmi},
  {Kretschmar}, {Hou}, \& {Takata}}]{li+17_0632}
{Li}, J., {Torres}, D.~F., {Cheng}, K.-S., {et~al.} 2017, \apj, 846, 169

\bibitem[{{Liu} {et~al.}(2006){Liu}, {van Paradijs}, \& {van den
  Heuvel}}]{lvv06}
{Liu}, Q.~Z., {van Paradijs}, J., \& {van den Heuvel}, E.~P.~J. 2006, VizieR
  Online Data Catalog, 345

\bibitem[{{Lutovinov} \& {Tsygankov}(2009)}]{lt09}
{Lutovinov}, A.~A., \& {Tsygankov}, S.~S. 2009, Astronomy Letters, 35, 433

\bibitem[{{Lyne} {et~al.}(2015){Lyne}, {Stappers}, {Keith}, {Ray}, {Kerr},
  {Camilo}, \& {Johnson}}]{lyn+15}
{Lyne}, A.~G., {Stappers}, B.~W., {Keith}, M.~J., {et~al.} 2015, \mnras, 451,
  581

\bibitem[{{Matsuoka} {et~al.}(2009){Matsuoka}, {Kawasaki}, {Ueno}, {Tomida},
  {Kohama}, {Suzuki}, {Adachi}, {Ishikawa}, {Mihara}, {Sugizaki}, {Isobe},
  {Nakagawa}, {Tsunemi}, {Miyata}, {Kawai}, {Kataoka}, {Morii}, {Yoshida},
  {Negoro}, {Nakajima}, {Ueda}, {Chujo}, {Yamaoka}, {Yamazaki}, {Nakahira},
  {You}, {Ishiwata}, {Miyoshi}, {Eguchi}, {Hiroi}, {Katayama}, \&
  {Ebisawa}}]{maxi09}
{Matsuoka}, M., {Kawasaki}, K., {Ueno}, S., {et~al.} 2009, \pasj, 61, 999

\bibitem[{{Meurs} \& {van den Heuvel}(1989)}]{mv89}
{Meurs}, E.~J.~A., \& {van den Heuvel}, E.~P.~J. 1989, \aap, 226, 88

\bibitem[{{Papitto} \& {Torres}(2015)}]{pt15}
{Papitto}, A., \& {Torres}, D.~F. 2015, \apj, 807, 33

\bibitem[{{Papitto} {et~al.}(2014){Papitto}, {Torres}, \& {Li}}]{pap+14}
{Papitto}, A., {Torres}, D.~F., \& {Li}, J. 2014, \mnras, 438, 2105


\bibitem[Rando \& for the Fermi LAT Collaboration(2009)]{ran+09} Rando, R., \& for the Fermi LAT Collaboration 2009, arXiv:0907.0626

\bibitem[{{Reig}(2011)}]{rei11}
{Reig}, P. 2011, \apss, 332, 1

\bibitem[Riquelme et al.(2012)]{rtn12} Riquelme, M.~S., Torrej{\'o}n, J.~M., \& Negueruela, I.\ 2012, \aap, 539, A114

\bibitem[{{Romero} {et~al.}(2001){Romero}, {Kaufman Bernad{\'o}}, {Combi}, \&
  {Torres}}]{rom+01}
{Romero}, G.~E., {Kaufman Bernad{\'o}}, M.~M., {Combi}, J.~A., \& {Torres},
  D.~F. 2001, \aap, 376, 599

\bibitem[{{Shrader} {et~al.}(1999){Shrader}, {Sutaria}, {Singh}, \&
  {Macomb}}]{shr+99}
{Shrader}, C.~R., {Sutaria}, F.~K., {Singh}, K.~P., \& {Macomb}, D.~J. 1999,
  \apj, 512, 920

\bibitem[{{Stappers} {et~al.}(2014){Stappers}, {Archibald}, {Hessels}, {Bassa},
  {Bogdanov}, {Janssen}, {Kaspi}, {Lyne}, {Patruno}, {Tendulkar}, {Hill}, \&
  {Glanzman}}]{sta+14}
{Stappers}, B.~W., {Archibald}, A.~M., {Hessels}, J.~W.~T., {et~al.} 2014,
  \apj, 790, 39

\bibitem[{{Stollberg} {et~al.}(1993){Stollberg}, {Finger}, {Wilson}, {Harmon},
  {Rubin}, {Zhang}, \& {Fishman}}]{sto+93}
{Stollberg}, M.~T., {Finger}, M.~H., {Wilson}, R.~B., {et~al.} 1993, \iaucirc,
  5836

\bibitem[{{Takata} {et~al.}(2014){Takata}, {Li}, {Leung}, {Kong}, {Tam}, {Hui},
  {Wu}, {Xing}, {Cao}, {Tang}, {Wang}, \& {Cheng}}]{tak+14}
{Takata}, J., {Li}, K.~L., {Leung}, G.~C.~K., {et~al.} 2014, \apj, 785, 131

\bibitem[{{Tam} {et~al.}(2011){Tam}, {Huang}, {Takata}, {Hui}, {Kong}, \&
  {Cheng}}]{tam+11}
{Tam}, P.~H.~T., {Huang}, R.~H.~H., {Takata}, J., {et~al.} 2011, \apjl, 736,
  L10

\bibitem[{{Tam} {et~al.}(2015){Tam}, {Li}, {Takata}, {Okazaki}, {Hui}, \&
  {Kong}}]{tam+15}
{Tam}, P.~H.~T., {Li}, K.~L., {Takata}, J., {et~al.} 2015, \apjl, 798, L26

\bibitem[{{Tsygankov} {et~al.}(2017){Tsygankov}, {Wijnands}, {Lutovinov},
  {Degenaar}, \& {Poutanen}}]{tsy+17}
{Tsygankov}, S.~S., {Wijnands}, R., {Lutovinov}, A.~A., {Degenaar}, N., \&
  {Poutanen}, J. 2017, \mnras, 470, 126

\bibitem[{{Walter} {et~al.}(2015){Walter}, {Lutovinov}, {Bozzo}, \&
  {Tsygankov}}]{wal+15}
{Walter}, R., {Lutovinov}, A.~A., {Bozzo}, E., \& {Tsygankov}, S.~S. 2015,
  \aapr, 23, 2

\bibitem[{{Wang} {et~al.}(2009){Wang}, {Archibald}, {Thorstensen}, {Kaspi},
  {Lorimer}, {Stairs}, \& {Ransom}}]{wan+09}
{Wang}, Z., {Archibald}, A.~M., {Thorstensen}, J.~R., {et~al.} 2009, \apj, 703,
  2017

\bibitem[{{Wang} {et~al.}(2013){Wang}, {Breton}, {Heinke}, {Deloye}, \&
  {Zhong}}]{wan+13}
{Wang}, Z., {Breton}, R.~P., {Heinke}, C.~O., {Deloye}, C.~J., \& {Zhong}, J.
  2013, \apj, 765, 151

\bibitem[{{Xing} \& {Wang}(2013)}]{xw13}
{Xing}, Y. \& {Wang}, Z. 2013, \apj, 769, 119

\bibitem[{{Xing} {et~al.}(2016){Xing}, {Wang}, \& {Takata}}]{xwt16}
{Xing}, Y., {Wang}, Z., \& {Takata}, J. 2016, \apj, 828, 61

%%\bibitem[{{Xing} {et~al.}(2017){Xing}, {Wang}, \& {Takata}}]{xwt17} {Xing}, Y., {Wang}, Z., \& {Takata}, J. 2016, \apj, 828, 61

\bibitem[{{Xing} {et~al.}(2018){Xing}, {Wang}, \& {Takata}}]{xwt18}
{Xing}, Y., {Wang}, Z.-X., \& {Takata}, J. 2018, Research in Astronomy and
  Astrophysics, 18, 127

\end{thebibliography}
\end{document}